\documentstyle [psfig]{l-aa}

\input psfig

\def\lesssim{\mathrel{\mathpalette\fun <}}
\def\fun#1#2{\lower3.6pt\vbox{\baselineskip0pt\lineskip.9pt
  \ialign{$\mathsurround=0pt#1\hfil##\hfil$\crcr#2\crcr\sim\crcr}}}

\begin{document}

   \thesaurus{
		12.03.4 
		02.07.2 
	      %
	      %
	      %
	      %
		} 

   \title{
		Determining the Cosmic Equation of State Using
		Future Gravitational Wave Detectors
	  }

 \author{Zong-Hong Zhu\inst{1,2} 
		\and Masa-Katsu Fujimoto\inst{1}
		\and Daisuke Tatsumi\inst{1} 
	}

  \offprints{Z.-H. Zhu (zong-hong.zhu@nao.ac.jp)}

 \institute{National Astronomical Observatory, 
			2-21-1, Osawa, Mitaka, Tokyo 181-8588, Japan
		\and 
			Beijing Astronomical Observatory,
			Chinese Academy of Sciences, Beijing 100012, China
	      }

  \date{Received 00 00, 0000; accepted 00 00, 0000}

  \maketitle

  \markboth{Z.-H. Zhu et al.: 
		Determining the cosmic equation of state using 
		future gravitational wave detectors}{}

\begin{abstract}

The expected chirp mass distribution of observed events for future
gravitational wave detectors is extensively investigated in the presence
of an exotic fluid component with an arbitrary equation of state,
$-1 \leq \omega_x \equiv p_x/\rho_x < 0$, i.e., the so-called
dark energy component.
The results for a flat model dominated by a dark energy are compared
to those for the standard flat model dominated by cold dark matter.
It is found that for a flat universe the chirp mass distribution
shows a sensitive dependence on $\omega_x$, which may provide an
independent and robust constraint on the cosmic equation of state.\\


\end{abstract}

\section{Introduction}

Upcoming interferometric detectors of gravitational waves such as
LIGO (Abramovici et al. 1992),
VIRGO (Caron et al. 1997),
GEO (Danzmann et al. 1997) and 
TAMA (Kuroda 1997)
will produce a flood of data when they come online in a few years.
Advanced detectors in LIGO/VIRGO and 
LCGT (Kuroda et al. 1999) will even be
able to see neutron star binaries (NS/NS) out to cosmological redshifts.
Because the accuracy in the measurement of the signal strength can be better 
than 10\%, and the accuracy in the measurement of the chirp mass 
(which characterizes the frequency sweep of a binary inspiral) can be better 
than 0.1\%(Finn \& Chernoff 1993; Cutler \& Flanagan 1994),
observations of NS/NS inspiral events can be used to determine the universe's
 Hubble constant, deceleration parameter and cosmological constant
 (Schutz 1986, 1989; Markovi\'{c} 1993; Chernoff \& Finn 1993; 
	Finn 1996; Wang \& Turner 1997).
In particular, Finn (1996) and Wang \& Turner (1997) explicitly pointed out
that the chirp mass spectrum of observed neutron star binary inspiral 
events can be quite sensitive to cosmological constant.

In this paper, 
we investigate the possibility of using the future
gravitational wave data to constrain dark energy, an exotic fluid component
in the universe with an arbitrary equation of state,
$-1 \leq \omega_x \equiv p_x/\rho_x < 0$,
whose existence has been well motivated by various astronomical observations 
 and theoretical arguments (refer to a recent review, Turner 1999).
While the cosmological constant (vacuum energy) can apparently succeed 
in explaining the accelerating expansion of the universe which
has stirred cosmologists after observations and measurements of distant
Ia type supernovae (SNeIa) (Perlmutter et al. 1999; Riess et al. 1998),
it suffers from the difficulty 
in understanding of the observed value in the framework of modern  quantum 
field theory (Weinberg 1989; Carroll et al. 1992) and
the ``coincidence problem'', the issue of 
explaining the initial conditions necessary to yield the near-coincidence 
of the densities of matter and the cosmological constant component today.
In this case, a dark energy component  with generally negative pressure
has been invoked.
Examples of dark energy include the evolving scalar field 
(Caldwell et al. 1998; Zlatev et al. 1999; Steinhardt et al. 1999),
the smooth  noninteracting component (Turner \& White 1997; Chiba et al. 1997)
and the frustrated network of topological defects in which 
$\omega_x = - \frac{n}{3}$,
with n being the dimension of the defect (Spergel \& Pen 1997).
Some observational constraints for these models have been extensively 
analyzed in the literature, 
including the redshift-luminosity distance relation based on distant 
      SNeIa (Garnavich et al. 1998),
      gravitational lensing statistics (Cooray \& Huterer 1999; 
					Zhu \& Cao 1999; Zhu 2000),
      Einstein ring system (Futamase \& Hamana 1999),
      the anisotropy of the Cosmic Microwave Background Radiation 
		(CMBR) (Huey et al. 1999),
      the angular size-redshift relation of distant extragalatic
                sources (Lima \& Alcaniz 2000a),
      the age of high-redshift galaxies (Lima \& Alcaniz 2000b),
and the combinations of SNeIa measurements with 
		either gravitational lensing (Waga \& Miceli 1999)
		or CMBR (Efstathiou 1999)
		or large-scale structure (Perlmutter et al. 1999)
		or the time variations of the redshift of quasars (Nakamura \&
								   Chiba 1999).
A combined maximum likelihood  analysis has suggested a range of 
 equation-of-state for quintessence, $-1 \le w_x \lesssim -0.6$ 
 (see Wang et al. 2000 for an elegant summary).

Because the determination of the amount and nature of dark energy is
 emerging as one of the most important challenges in cosmology,
 it is worthwhile to investigate other observational aspects of its
 equation of state.
In this work, from calculating the expected chirp mass spectrum of observed 
 NS/NS inspiral events for future gravitational wave detectors in the presence
 of a dark energy component, we find a sensitive dependence of chirp mass 
 distribution on $\omega_x$ for a flat universe, which could provide an
 independent and robust constraint on the cosmic equation of state.
For calculations of this work, our notation and approach mainly follow
 those described by Finn (1996).

\section{The chirp mass spectrum of NS/NS inspiral events}

The NS/NS inspirals are the targets for future gravitational wave detectors,
whose merger rate at redshift $z$ per unit observer time interval
per unit physical volume is, $\dot{n} = \dot{n}_0 \, F(z)\,(1+z)^2$,
where $F(z)$ describes the NS/NS merge rate evolution, 
and $\dot{n}_0 (\simeq 10^{-7} h\, {\rm Mpc}^{-3} {\rm yr}^{-1})$
is the local NS/NS merger rate per unit volume (Phinney 1991).
In the case of no proper evolution $F(z)=1$.
The evolution of the NS/NS merger rate has been studied by several
authors (e.g., Tutukov \& Yungelson 1993; Lipunov et al. 1995).
Because recent observations of field galaxies out to redshift $z\sim 5$
(see e.g. Madau et al. 1998; Steidel et al. 1999)
have provided a means of modeling the evolution of the star formation
rate (SFR), one can now compute the birth and merger rates of binary systems
by taking into account the time-delay. Using population synthesis model,
Schneider et al. (2000) have calculated birth and merger rate evolution
of various kinds of binaries. Our best fit to their results for NS/NS
merger rate from hierarchical scenario (Fig.8 of Schneider et al. 2000) takes
\begin{equation}
\label{evolution}
F(z) = 1 + 11.5z^2 - 13.6z^3 + 5.37z^4 - 0.70z^5 \,\,\,\,\,\,\,\, z\in (0,3).
\end{equation}

For a NS/NS inspiral source located at redshift $z$, the detectors
measure its chirp mass,
${\cal M} \equiv {\cal M}_0 \, (1+z)$
where the intrinsic chirp mass,
${\cal M}_0 \equiv \mu^{3/5} M^{2/5}$,
with $\mu$ and $M$ being the binary's reduced and total mass respectively.
Therefore, the chirp mass spectrum represents directly the source
 redshift distribution if ${\cal M}_0$ is constant.
Because we are considering NS/NS binaries, ${\cal M}_0=1.19\,{\rm M}_{\odot}$
 will be used below, corresponding to a typical neutron star mass
 $m=1.37\,{\rm M}_{\odot}$.
The signal-to-noise ratio $\rho$ of an event observed by a given
 detector is 
$
\rho= 8 \Theta\, (r_0/D_L)\left(  {\cal M}/1.2 {\rm M}_{\odot}\right)
^{5/6} \, \zeta(f_{\rm max}),
$
where $D_L$ is the luminosity distance from the detector to the source.
$\zeta$($\simeq 1$ for our case) describes the overlap of
the signal power with the detector bandwidth, while $r_0$ gives an
overall sense of the depth to which the detector can reach.
The signal-to-noise ratio of an inspiraling binary with a single
interferometric detector depends on the relative orientation of the
source, which is confined to the angular orientation function $\Theta$.
From observations of binary inspiral in a single interferometer
we can measure $\rho$ and $\cal M$ but not $\Theta$.
Fortunately, even though $\Theta$ cannot be measured,
its probability distribution can be approximated by (Finn 1996)
\begin{equation}
P_\Theta(\Theta) = \left\{
\begin{array}{ll}
5\Theta\left(4-\Theta\right)^3/256 &  {\rm ~~if~~} 0<\Theta<4\\
0 & {\rm ~~otherwise.}
\end{array}\right.
\end{equation}
which is related to $P_\rho(\rho|{\cal M}_0,{\cal M})$,
the probability distribution of signal-to-noise ratio
of events with chirp mass ${\cal M}$, through
\begin{eqnarray}
\label{Proh}
P_\rho(\rho|{\cal M}_0,{\cal M}) & = & 
	P_\Theta\left[{\rho\over8}{D_L(Z)\over r_0}
        \left(1.2\,{\rm M}_\odot\over{\cal M}\right)^{5/6}\right] 
        \nonumber\\
 & &  \times {D_L(Z)\over8 r_0}
        \left(1.2\,{\rm M}_\odot\over{\cal M}\right)^{5/6},
\end{eqnarray}
where $Z={\cal M} / {\cal M}_0 -1$.
Then, in a homogeneous and isotropic universe we get the rate at which binary
inspiral signals corresponding to chirp mass $\cal M$ and $\rho$
greater than $\rho_0$ are detected
\begin{eqnarray}
\label{dNdot}
\frac{{\rm d}\dot{N}(>\rho_0, {\cal M})}{{\rm d} ({\cal M}/{\cal M}_0)} &  = &
    \dot{n}\, \cdot 4\pi D_A^2 \left.\frac{ c\,{\rm d} t}{{\rm d} z}\right|_Z
    \cdot \int_{\rho_0}^\infty P_\rho(\rho|{\cal M}_0,{\cal M}) {\rm d}  \rho
    \nonumber\\
\frac{}{} & = & 4\pi \dot{n}_0\, F(Z)\, \left( \frac{D_L(Z)}{1+Z}\right)^2
     \left.\frac{ c\,{\rm d} t}{{\rm d} z}\right|_Z
    \nonumber\\ 
     & & \times C_\Theta\left[ {\rho_0\over8}{D_L(Z)\over r_0}
     \left(1.2\,{\rm M}_\odot\over{\cal M}\right)^{5/6} \right] ,
\end{eqnarray}
where $D_A$ is the angular diameter distance from the detector to the source,
$C_{\Theta}(x)$ is the probability that a given detector receives
a binary inspiral with chirp mass ${\cal M}$ and signal-to-noise ratio greater
than $\rho_0$:
\begin{eqnarray}
C_{\Theta}(x) &= &\left\{ \begin{array}{ll}
(1+x)\,(4-x)^4 /256, \hskip 0.5cm  &\mbox{if $0<x<4$}, \\
0,      \hskip 1.5cm            &\mbox{otherwise}. \end{array} \right.
\end{eqnarray}
An accumulation of observed NS/NS binary inspiral events will give a
chirp mass spectrum
\begin{equation}
P({\cal M}|\rho_0) = \frac{{\rm d}\dot{N}(>\rho_0, {\cal M})
                        / {\rm d} {\cal M}}{\dot{N}(>\rho_0)}  \,\,\,\,\,.
\label{CMS}
\end{equation}

\begin{figure*}[t]
\hbox{
\psfig{figure=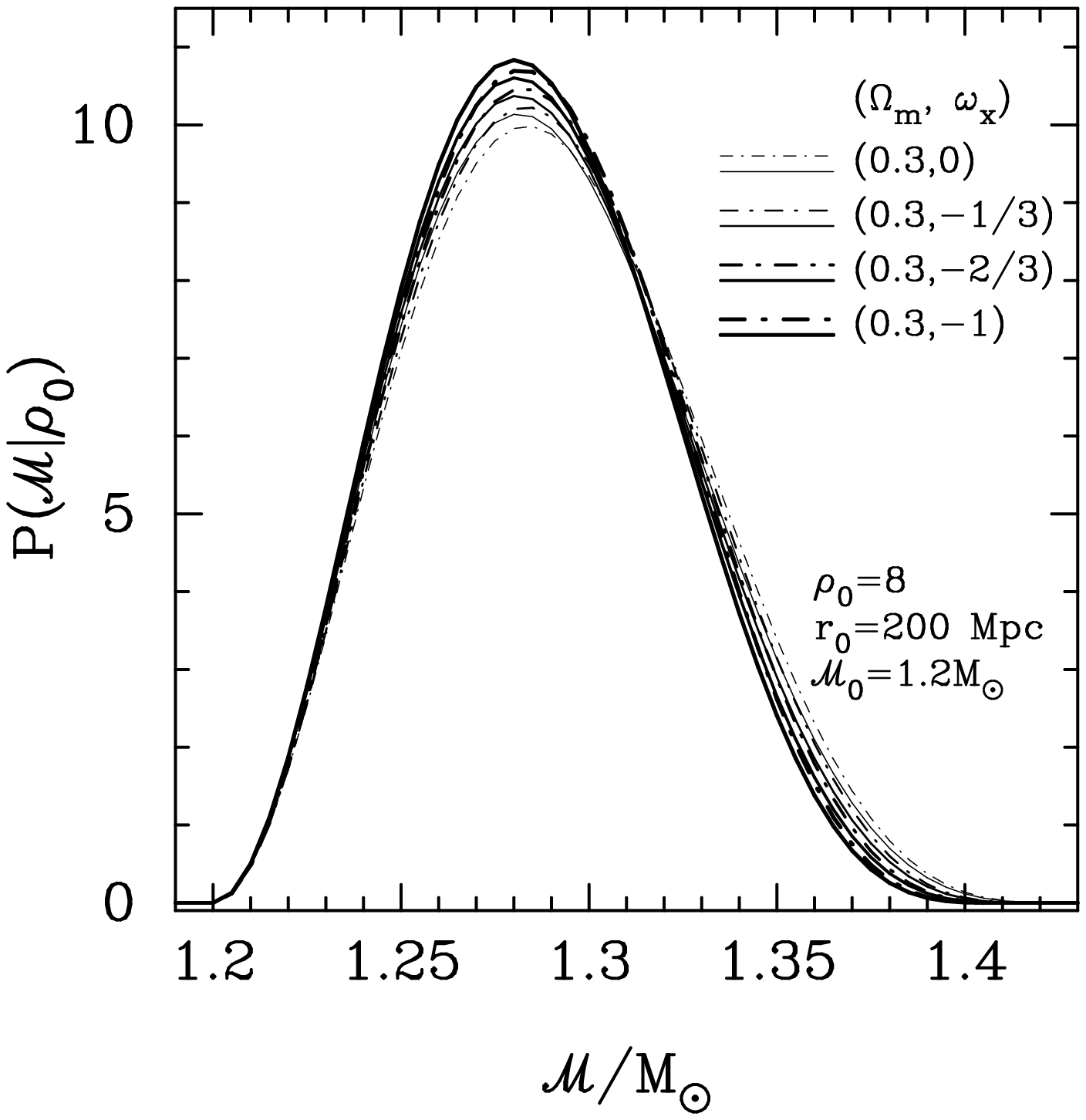,width=8.20cm}
\hspace*{1.0cm}
\hfill
\psfig{figure=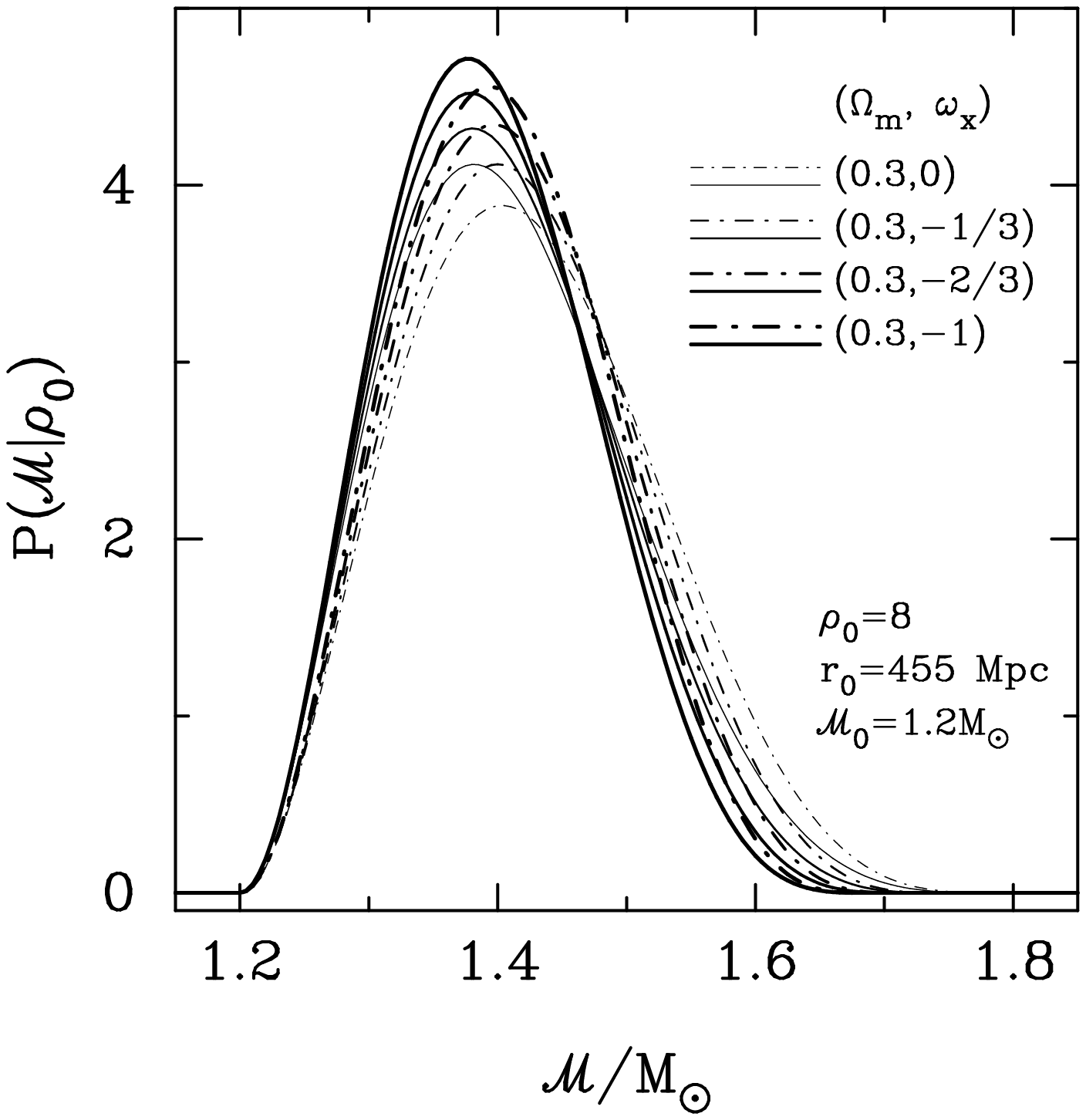,width=8.20cm}
}
\caption{
	The chirp mass spectrum of NS/NS binary inspirals for a flat universe,
        $(\Omega_m,\Omega_x) = (0.3,0.7)$,
        with $\omega_x=0$ (SCDM model), $-1/3$, $-2/3$ and
                $-1$ ($\Lambda$CDM model).
	The detector depth is assumed to be $r_0=200\,$Mpc (left panel)
	and $r_0=455\,$Mpc (right panel) respectively.
	Solid lines are for the nonevolution case while dash-dot lines are
	for the evolutionary case from Schneider et al. (2000).
        }
\label{spectrum-1m}
\end{figure*}

\begin{figure*}[t]
\hbox{
\psfig{figure=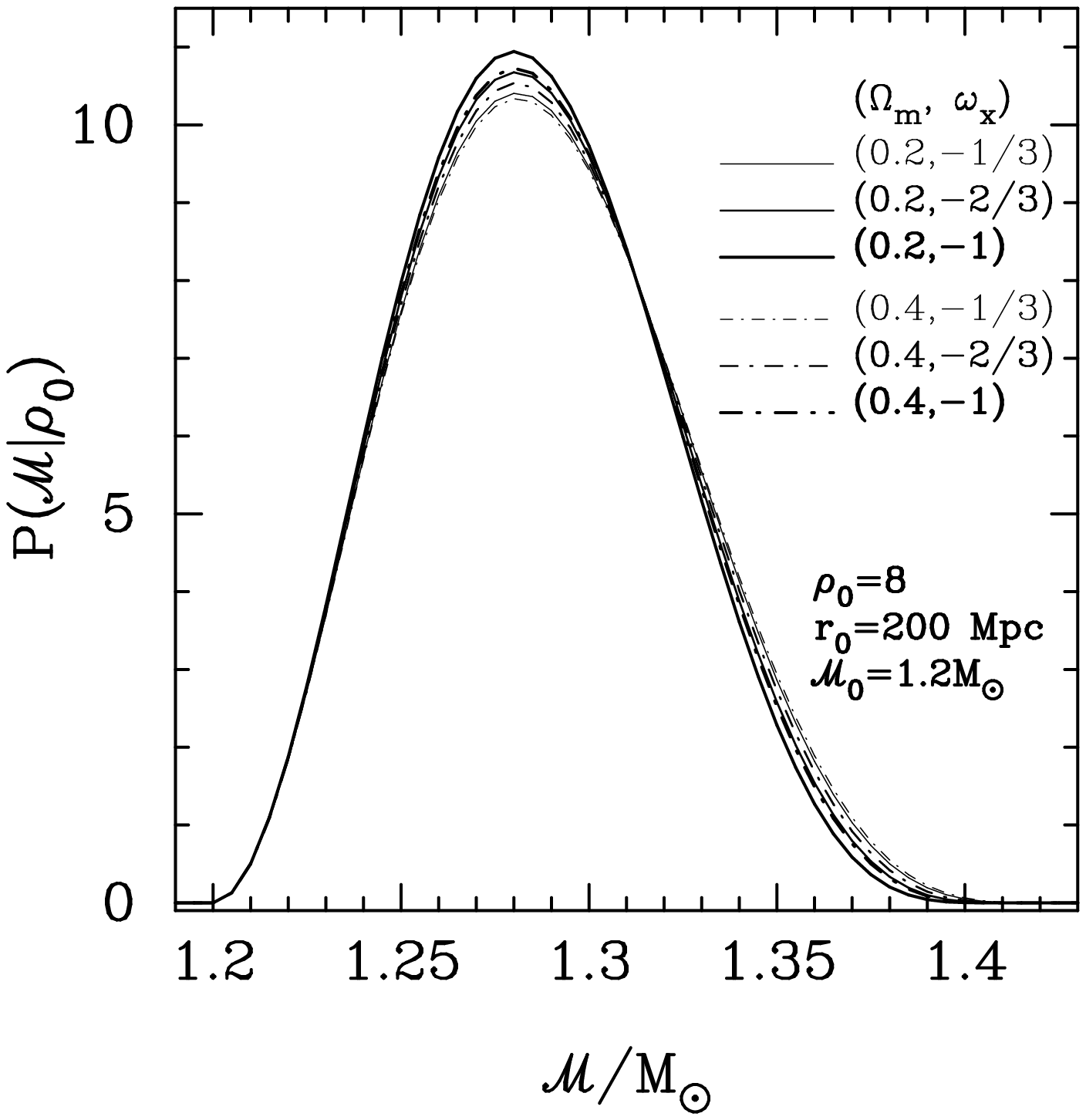,width=8.20cm}
\hspace*{1.0cm}
\hfill
\psfig{figure=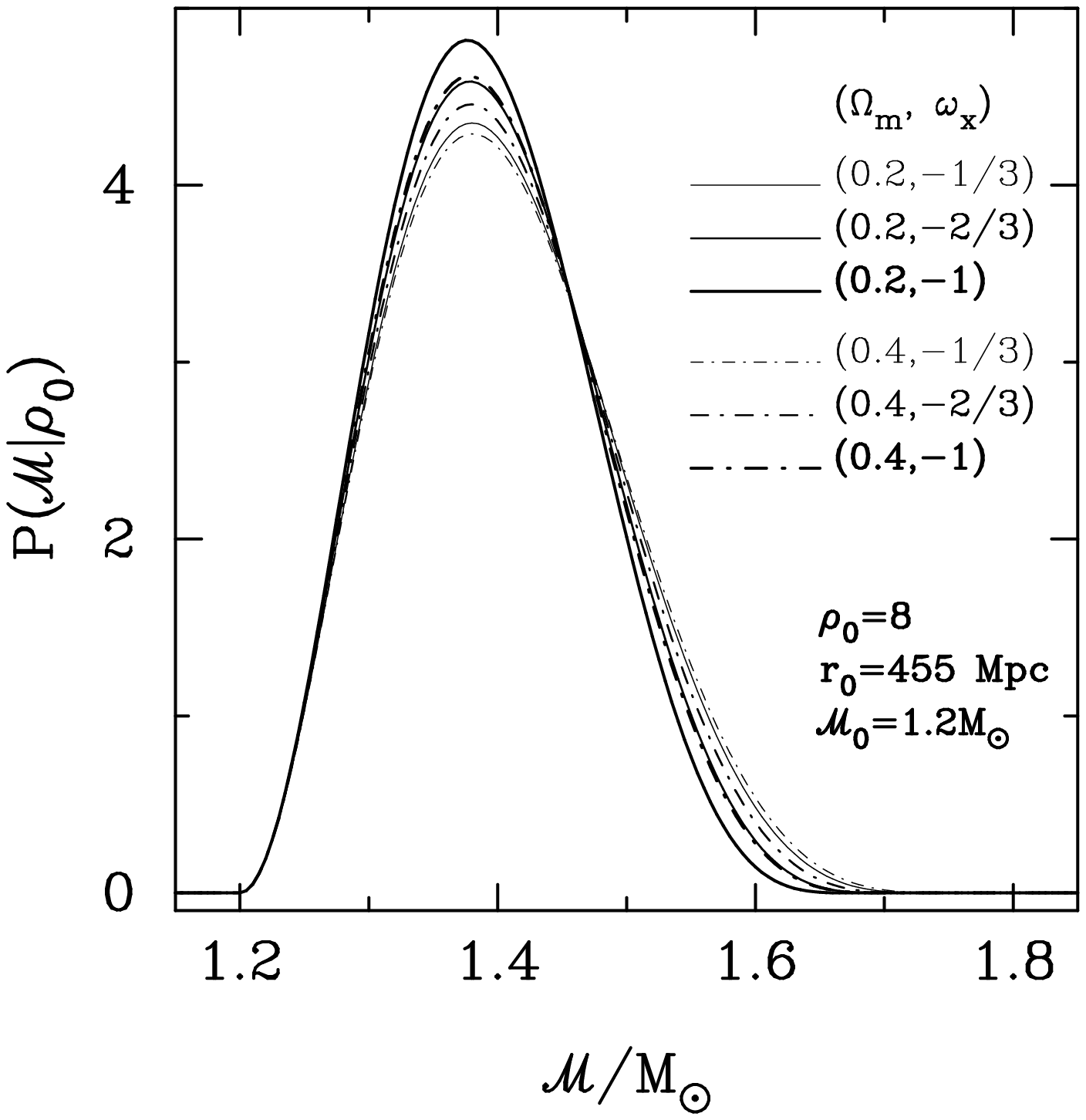,width=8.20cm}
}
\caption{
        The chirp mass spectrum of NS/NS binary inspirals for a flat universe
	with different matter density $\Omega_m=0.2$ (solid lines) and
	$\Omega_m=0.4$ (dash-dot lines).
        The detector depth is assumed to be $r_0=200\,$Mpc (left panel)
        and $r_0=455\,$Mpc (right panel) respectively.
        }
\label{spectrum-2m}
\end{figure*}

\section{Determining the cosmic equation of state}

In order to explicitly show the dependence of the chirp mass spectrum on
the cosmic equation of state, one needs to consider it in the context of general
Friedmann-Robertson-Walker (FRW) cosmologies in the presence of a dark energy 
component. For general FRW cosmologies, the metric of space time
is described by (in the $c = 1$ unit):
\begin{equation}
ds^2 = -dt^2 + R^2(t) \left[ d\chi^2 +
f^2(\chi)(d \theta^2 + \sin^2\theta d\phi^2)\right],
\label{metric}
\end{equation}
where $f(\chi) = \chi$ for a flat universe ($k=0$), $f(\chi) = \sin\chi$
for a closed universe ($k= + 1$), and $f(\chi) = \sinh\chi$ for an open
universe ($k= - 1$).
The comoving distance $\chi$ is (Zhu 1998, Zhu \& Cao 1999)
\begin{equation}
\chi = \left\{ \begin{array}{ll}
                \int_0^z \frac{dz}{\sqrt{\sum_i \Omega_i (1+z)^{3(1+\omega_i)}}}                                &       (k=0)\\
                & \\
                \left| \Omega_k \right|^{1/2} \int_0^{z}
                         \frac{dz}{\sqrt{\sum_{i} \Omega_i (1+z)^{3(1+\omega_i)}                        + \Omega_k (1+z)^2}}           & (k=\pm 1)\\
                \end{array}
        \right.
\label{comoving}
\end{equation}
where $i$ includes each component of matter and energy in the universe,
 $\omega_i = p_i/\rho_i$ parameterizes the effective equation of state
 for the $i$th-component, $\Omega_i \equiv ({8 \pi G }/{3 H_0^2})\rho_{i0}$,
 and $\Omega_k \equiv {-k}/{R_0^2 H_0^2}$ with $H_0$ being the Hubble constant
 ($H_0=100h\,$km$\,$sec$^{-1}$Mpc$^{-1}$, $0.5 \leq h <1$)
 and zero subscripts referring to the present epoch. 
Various distance measures, 
 including $D_L$ and $cdt/dz$ that are essential to calculate the chirp mass 
 spectrum Eq.(\ref{CMS})
 can be easily inferred from
 the above comoving distance (Zhu 1998, Zhu \& Cao 1999).

Fig.~\ref{spectrum-1m} explicitly demonstrates how the chirp mass spectrum 
depends sensitively on the equation of state $\omega_x$ with both cases
of NS/NS merger rate nonevolution (solid lines) and evolution (dash-dot lines).
We have use two detector depths, $r_0=200\,$Mpc for LCGT (left panel) and 
$455\,$Mpc for LIGOII\footnote{http://www.ligo.caltech.edu/~ligo2/index.html\\
\hspace*{0.4cm}
http://www.ligo.caltech.edu/docs/T/T990080-00.pdf} (right panel), respectively.
In our calculations, we use a flat universe:
$(\Omega_m,\Omega_x) = (0.3,0.7)$ with a constant $\omega_x \in (-1,0)$,
the dimensionless Hubble constant $h=0.65$
 and the signal-to-noise ratio threshold $\rho_0=8$.
As it is shown, there exists a sensitive dependence of the chirp mass
spectrum of NS/NS binary inspiral on the cosmic equation-of-state.
In general the smaller $\omega_x$, the more compressed the spectrum,
        and the smaller the tail at large ${\cal M}$.
This fact may provide an independent and robust constraint on the cosmic
 equation of state.

\section{Discussion}

The solid lines of Fig.~\ref{spectrum-1m} are for the case of NS/NS merger 
rate nonevolution, $F(z)=1$. 
Unfortunately the evolution rate is probably much strong.
The dash-dot lines of Fig.~\ref{spectrum-1m} are based on the result of
Schneider et al. (2000) (Eq.(\ref{evolution})). As is shown, the merger rate
evolution makes the spectrum less compressed and extends the tail towards
large ${\cal M}$, which is exactly opposite to the effect for lowering
$\omega_x$. Because uncertainties in the merger rate evolution directly lead to
uncertainties in determining the cosmological equation of state, the contraints
on dark energy from gravitational wave detectors could not be robust until
we understand well the cosmic evolution of the NS/NS binary merger rate.

Another uncertainty of the method proposed here comes from the degenerate
between the cosmic equation of state, $\omega_x$, and the cosmic matter 
density, $\Omega_m$. Fig.~\ref{spectrum-2m} explicitly shows how $\Omega_m$ 
changes the spectrum. A spectrum for a flat universe with lower $\Omega_m$ and
larger $\omega_x$ may overlap with the one for a flat universe with higher
$\Omega_m$ and smaller $\omega_x$. One need an independent cosmological test
to break down this degenerate, such as clusters of galaxies, gravitational 
lensing or the anisotropy of CMBR etc.

Last but not least, neutron star masses are not all identical
($1.37\,{\rm M}_{\odot}$), but distribute
between a lower bound (e.g. $1.29\,{\rm M}_{\odot}$) and a upper bound
(e.g. $1.45\,{\rm M}_{\odot}$).
This would introduce another probability function into Eqs.(\ref{Proh}) \&
(\ref{dNdot}).
Using a uniform distribution model, Finn (1996) shows that, the chirp mass
spectrum will broaden symmetrically as the mass distribution broadens.
Although the effect is different from the cosmic equation of state,
the uncertainty of neutron star mass distribution will also lead 
the uncertainty into the constraints on dark energy from future gravitational 
wave events catalogue.

As is shown above, the constraints on the cosmic equation of state from the 
 chirp mass spectrum of NS/NS merger events is promising, but it needs to 
 appeal to statisitics, evolution and neutron star mass distributions. 
It is worth mentioning that, if NS/NS merger events are observed to be 
 associated with gamma-ray bursts, one could measure the redshift of these 
 events through the galaxies hosting the associated GRBs. 
Then a few  NS/NS merger events would determine the cosmic equation of state 
 as what the high-$z$ SNIa has done.

\begin{acknowledgements}
We would like to thank Prof. E. S. Phinney for many helpful suggestions,
Prof. K. Kuroda for providing us the proposal materials for 
LCGT ({\em Large-scale Cryogenic Gravitational Wave Telescope}),
Prof. D. G. Blair and Dr. D. M. Coward for helpful discussion about the
 cosmic evolution of NS/NS inspiral event rate.
Z.-H.Zhu gratefully acknowledges support from a JSPS (the Japan Society for
the Promotion of Science) fellowship.
\end{acknowledgements}

\end{document}